\documentclass[3p,times]{elsarticle}
\usepackage{ecrc}

\volume{00}

\firstpage{1}

\journalname{Annals of Physics}

\runauth{S. Longhi}

\usepackage{amssymb}
\usepackage[figuresright]{rotating}

\begin{document}

\begin{frontmatter}



\title{Coherent transfer by adiabatic passage in two-dimensional lattices}


\author{Stefano Longhi}
\address{Dipartimento di Fisica, Politecnico di Milano and Istituto di Fotonica e Nanotecnologie del Consiglio Nazionale delle Ricerche, Piazza L. da Vinci 32, I-20133 Milano, Italy\\ Tel/Fax: 0039 022399 6156/6126, email: longhi@fisi.polimi.it}

\begin{abstract}
Coherent tunneling by adiabatic passage (CTAP) is a well-established technique for robust spatial transport of quantum particles in linear chains. 
Here we introduce two exactly-solvable models where the CTAP protocol can be extended to two-dimensional lattice geometries. Such bi-dimensional lattice models are synthesized from time-dependent second-quantization Hamiltonians, in which  the bosonic field operators evolve adiabatically like in an ordinary three-level CTAP scheme thus ensuring adiabatic passage in Fock space. 
\end{abstract}


\end{frontmatter}

\section{Introduction}

Adiabatic passage methods are interesting tools for manipulating quantum states of matter, which have found important application for a range of tasks including the manipulation of populations in atomic and molecular systems \cite{A1,A2,A3}, control of chemical reactions \cite{A3,A4},
 coherent quantum state transport \cite{A5,A6,A7}, and atomtronics \cite{A7bis}. Among other methods. stimulated Raman adiabatic
passage (STIRAP) \cite{A1,A2} is a robust technique for coherent
population transfer in a three-level quantum system, in which the population is transferred
adiabatically between two internal quantum states of an atom
by maintaining the system in a dark state. The spatial analogue of STIRAP,
referred to as coherent tunneling by adiabatic passage (CTAP), has been independently
proposed one decade ago  by two groups as a robust tool for quantum state transport of neutral atoms in optical traps \cite{A5} and of electrons
in quantum dot systems \cite{A6}. In CTAP, the quantum particle is transferred among positional quantum states by slowly changing
the tunneling interaction between the nearest neighboring
quantum units. Since the pioneering proposals of Refs.\cite{A5,A6},
spatial state transfer based on CTAP or similar adiabatic passage protocols has been 
extensively investigated by several authors \cite{A7,A7bis,A8,A9,A9bis,A10,A11}. CTAP has been proposed for creating a
maximally coherent superposition in a two-state atom \cite{A9}, for
manipulating single atoms in optical lattice \cite{A10}, electron spin
states in two-dimensional architecture \cite{A11}, and Bose-Einstein
condensates \cite{A12}. Optical analogues of CTAP have been proposed and 
experimentally demonstrated as an interesting tool
for robust light transfer among evanescently coupled optical waveguides \cite{A13,A14,A15} and for 
weakly-radiative wireless energy transfer \cite{Jann}. Spatial adiabatic passage processes for sound waves propagation in sonic crystals
have been proposed in Ref.\cite{A15bis}, whereas digital  CTAP, where the tunnel matrix elements are varied digitally
rather than smoothly, has been studied in Ref.\cite{A16}.
Finally, CTAP-based protocols have been applied to the problem
of transferring quantum states across a lattice  for realizing
long-range quantum state transfer \cite{A17}, quantum fan-out \cite{A18}, electron \cite{A19}
and atomic \cite{A20} interferometry. In spite of such an amount of works, there are very few studies on the possibility to apply CTAP or other spatial adiabatic passage methods to complex or bi-dimensional lattice geometries \cite{A11,A20,A20bis,A20tris,multi}. In particular, adiabatic spatial passage of a single cold atom in a system of three identical harmonic traps in a triangular geometry was investigated in Ref.\cite{A20}, and the different behavior arising from the two dimensional geometry as compared to the linear CTAP scheme was highlighted. In Ref.\cite{A20tris}, the Bose-Hubbard Hamiltonian was applied to a three-well system and it was shown analytically that CTAP of $N$ non-interacting particles across the chain is possible. Interestingly, in the Hilbert space such an Hamiltonian realizes CTAP in a bi-dimensional half-square lattice. \par
In this work we introduce exactly-solvable bi-dimensional lattice models, in which CTAP between distant sites can be realized  under appropriate dynamical control of the tunneling rates. In such models, the single-particle hopping dynamics on a bi-dimensional lattice is effectively reduced to that of a standard three-level STIRAP scheme on a linear chain by mapping the dynamical problem into  integrable second-quantization Hamiltonian models which are quadratic in the bosonic field operators. In the second-quantization framework, the bosonic field operators evolve adiabatically like in an ordinary three-level CTAP scheme, thus ensuring adiabatic passage in Fock space. An example of an exactly-solvable second-quantization model leading to an effective bi-dimensional particle dynamics in Fock space was recently presented by Bradly and coworkers in Ref.\cite{A20tris}. Though the motivation of that work was mainly concerned with the analytic solution to a Bose-Hubbard transport model, the underlying analysis showed that for $N$ non-interacting particles the dynamics in Hilbert space basically realizes CTAP in a half $N \times N$ square lattice, cut along one of the main diagonals.  Here we extend the second-quantization approach to realize bi-dimensional CTAP protocols by considering two different lattice geometries, namely rectangular and triangular lattices.

\section{CTAP in rectangular lattices}
One of the simplest extension of the linear-chain CTAP protocols to a bi-dimensional lattice is obtained by considering a rectangular lattice made of $N \times M$ sites, with $N$ and $M$ being arbitrary odd integers. In this lattice, for a general class of hopping rates the particle dynamics in the horizontal and vertical directions can be separated, and bi-dimensional CTAP can be thus realized by application of two independent multi-level CTAP schemes \cite{multiSTIRAP} along the horizontal ($N$ level scheme) and vertical ($M$ level scheme) directions. A different approach to  show CTAP in a rectangular lattice is to view the hopping motion of the particle in the bi-dimensional lattice as the dynamics in Fock space of a second-quantization Hamiltonian for certain bosonic fields. Although for the rectangular lattice problem such an approach  is more involved than the simpler separation of variable method, it is far more general since it can be extended to situations where separation of variables in not possible (like for the triangular lattice problem discussed in the next section). For the sake of clearness, we will consider here the case of a $3 \times 3$ lattice [see Fig.1(a)], though the analysis could be extended to a general $N \times M$ rectangular lattice, with $N$ and $M$ odd numbers. In the tight-binding and nearest-neighbor approximations, the amplitude probabilities $c_l(t)$ to find the particle at site $l$ of the rectangular lattice of Fig.1(a) ($l=1,2,3,...,9$) evolve according to the coupled equations
\begin{eqnarray}
i \frac{dc_1}{dt} & = & \Omega_4 c_2+\Omega_1 c_4 \\
i \frac{dc_2}{dt} & = & \Omega_4 c_1+\Omega_2 c_3+\Omega_1 c_ 5 \\
i \frac{dc_3}{dt} & = & \Omega_2 c_2+\Omega_1 c_ 6 \\
i \frac{dc_4}{dt} & = & \Omega_1 c_1+\Omega_4 c_5+\Omega_3 c_ 7 \\
i \frac{dc_5}{dt} & = & \Omega_1 c_2+\Omega_4 c_ 4+\Omega_2 c_6+\Omega_3 c_8 \\
i \frac{dc_7}{dt} & = & \Omega_3 c_4+\Omega_4 c_8 \\
i \frac{dc_6}{dt} & = & \Omega_1 c_3+\Omega_2 c_5+\Omega_3 c_ 9 \\
i \frac{dc_8}{dt} & = & \Omega_3 c_5+\Omega_4 c_ 7+\Omega_2 c_9 \\
i \frac{dc_9}{dt} & = & \Omega_3 c_6+\Omega_2 c_8 
\end{eqnarray}
where $\Omega_l=\Omega_l(t)$ ($l=1,2,3,4$) are the hopping rates between adjacent sites, as shown in Fig.1(a). We want to show that, under appropriate dynamical tuning of the hopping rates, perfect state transfer between the vertex states $|3 \rangle$ and $|7 \rangle$ (or, similarly, between the states $|1 \rangle$ and $|9 \rangle$ ) can be realized. Such a system thus provides a simple yet nontrivial extension, to a two-dimensional lattice, of the STIRAP method for a three-level system. A direct proof of the above mentioned property can be given by noticing that the system of Eqs.(1-9) admits of an instantaneous {\it dark} state with vanishing energy, given by
\begin{equation}
c_1=-\frac{\Omega_2 \Omega_3}{\sqrt{(\Omega_2^2+\Omega_4^2)(\Omega_1^2+\Omega_3^2)}} , \;\;  c_3=\frac{\Omega_3 \Omega_4}{\sqrt{(\Omega_2^2+\Omega_4^2)(\Omega_1^2+\Omega_3^2)}} ,\; \; 
c_7=\frac{\Omega_1 \Omega_2}{\sqrt{(\Omega_2^2+\Omega_4^2)(\Omega_1^2+\Omega_3^2)}}, \; \;  c_9=-\frac{\Omega_1 \Omega_4}{\sqrt{(\Omega_2^2+\Omega_4^2)(\Omega_1^2+\Omega_3^2)}} 
\end{equation}

\begin{figure}[b]
\includegraphics[width=12cm]{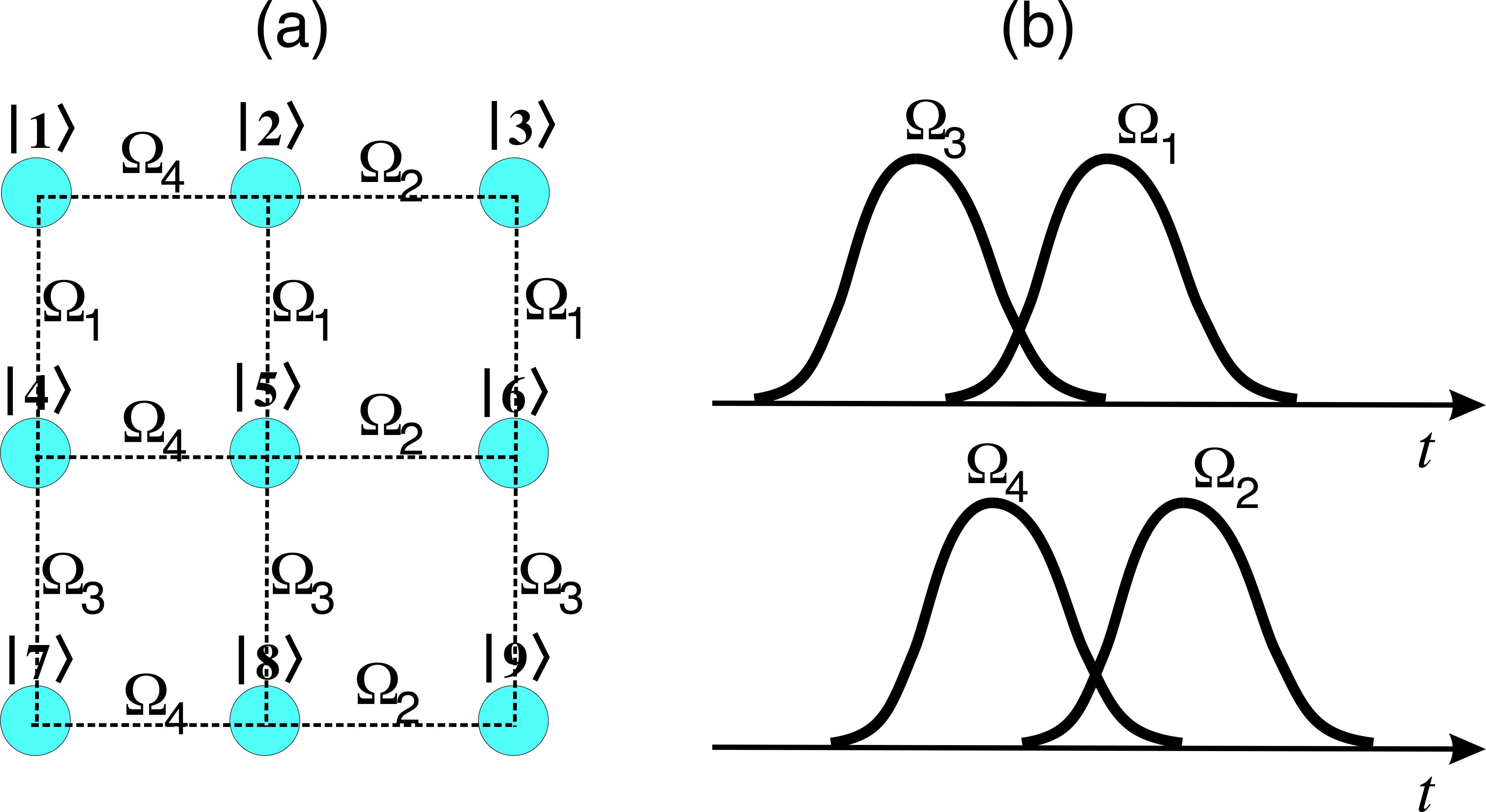}
\caption{(Color online). (a) Schematic of a square lattice made of $ 3 \times 3$ sites with hopping rates determined by the four parameters $\Omega_1(t)$, $\Omega_2(t)$, $\Omega_3(t)$ and $\Omega_4(t)$. (b) Typical sequence of the hopping rates that realizes  CTAP from the site $|3 \rangle$ to the site $|7 \rangle$.} 
\end{figure}

\begin{figure}[b]
\includegraphics[width=16cm]{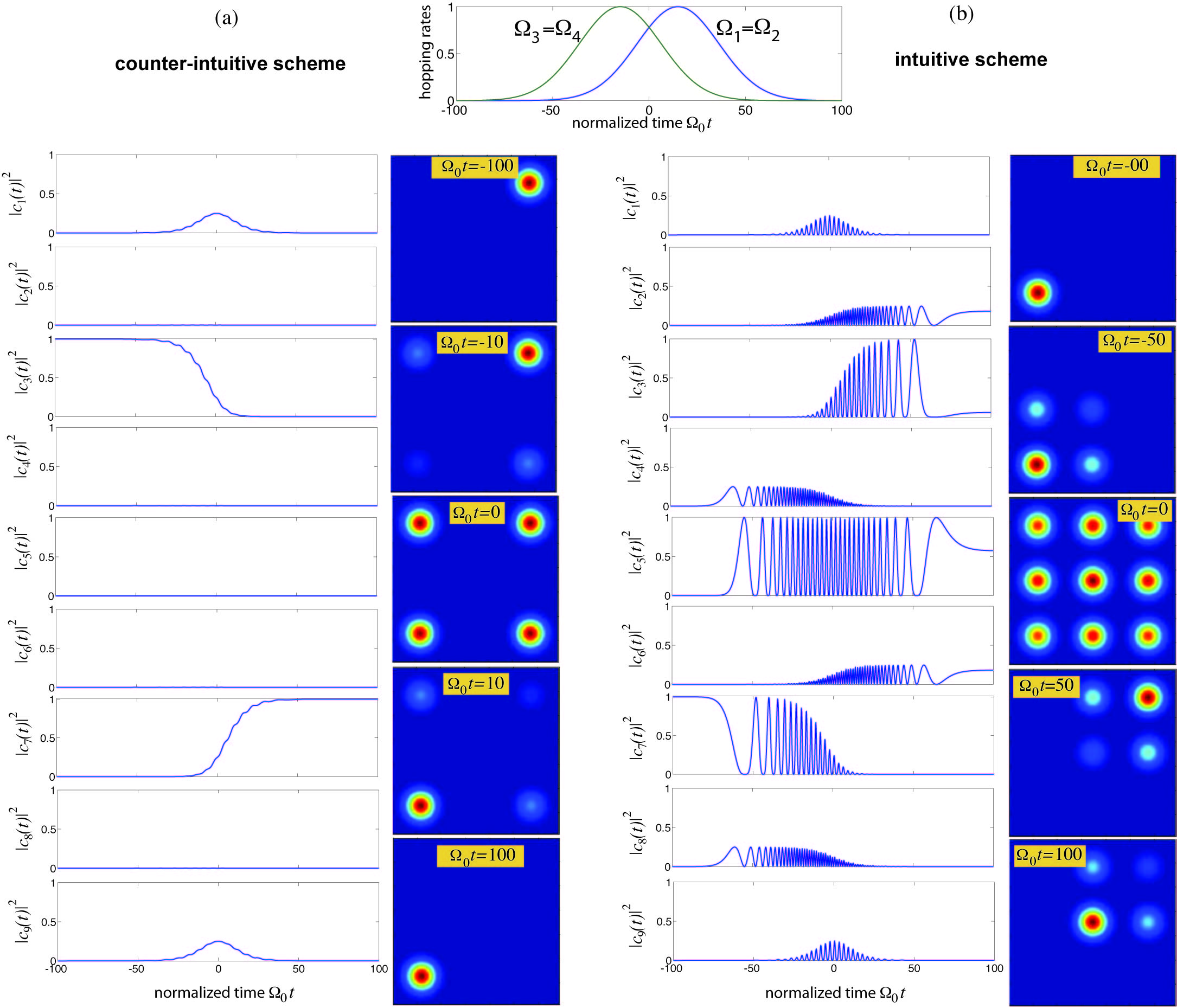}
\caption{(Color online). Numerically-computed evolution of populations (left panels) and snapshots of site occupation at a few times (right panels) in the $3 \times 3$ lattice of Fig.1(a) for initial excitation of (a) site $|3 \rangle$, and (b) site $|7 \rangle$. The 
 hopping rates $\Omega_l(t)$ used in the simulations, in units of $\Omega_0$, are shown in the upper inset.} 
\end{figure}

\begin{figure}[b]
\includegraphics[width=16cm]{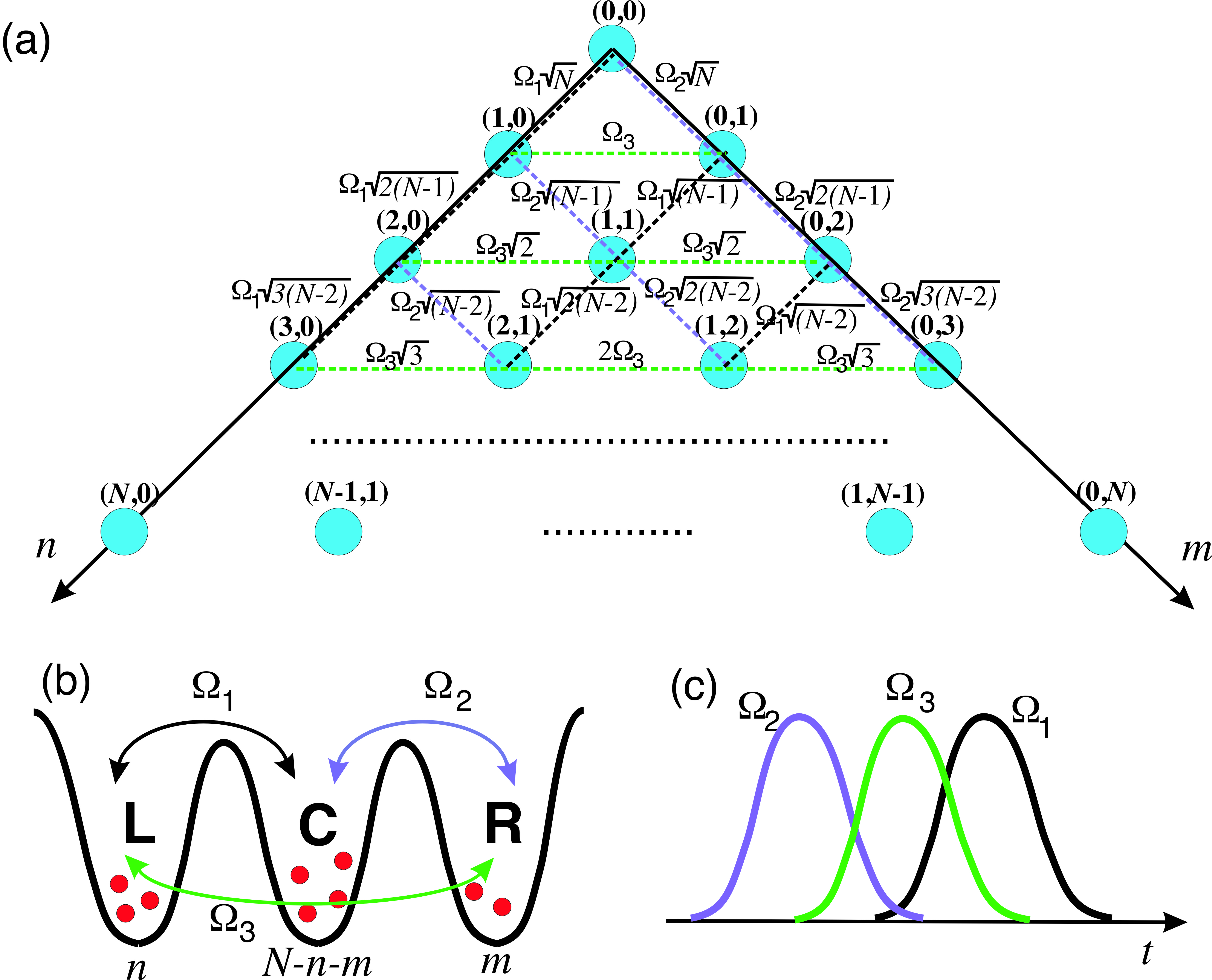}
\caption{(Color online). (a) Schematic of a triangular lattice composed by $(N+1)(N+2)/2$ sites with inhomogeneous hopping rates that depend on the three parameters $\Omega_1(t)$, $\Omega_2(t)$ and $\Omega_3(t)$. (b) Tunneling of $N$ non-interacting bosonic particles in a triple well potential with next-nearest hopping, that realizes in Fock space the triangular lattice model in (a). (c) Typical sequence of the hopping amplitudes $\Omega_1$, $\Omega_2$ and $\Omega_3$ that realizes CTAP from the state $(N,0)$ (all the bosons are initially in the left well) to the state $(N,0)$ (all the bosons are finally in the right well). For $\Omega_3=0$ the lattice in (a) describes a half-square lattice.} 
\end{figure}

\begin{figure}[b]
\includegraphics[width=16cm]{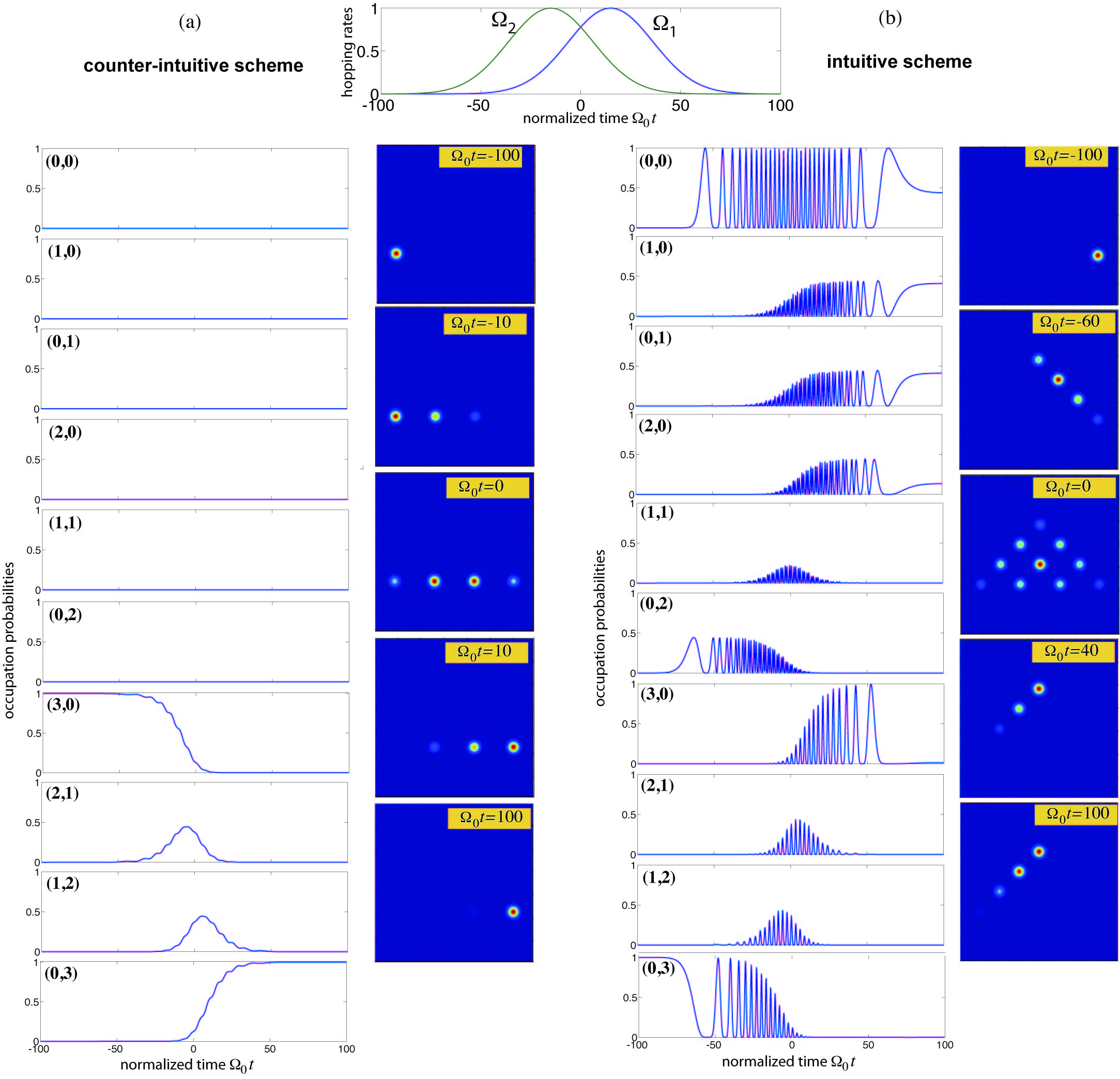}
\caption{(Color online). Numerically-computed evolution of populations $|c_{n,m}(t)|^2$ (left panels) and snapshots of site occupation at a few times (right panels) in  the half-square lattice of Fig.3(a) with $N=3$, $\Omega_3=0$ for initial excitation of (a) site $(3,0)$ [the counter-intuitive scheme], and (b) site $(0,3)$ [the intuitive scheme]. The upper inset shows the dynamical evolution of the tunneling rates $\Omega_1(t)$ and $ \Omega_2(t)$, in units of $\Omega_0$.} 
\end{figure}

\begin{figure}[b]
\includegraphics[width=13cm]{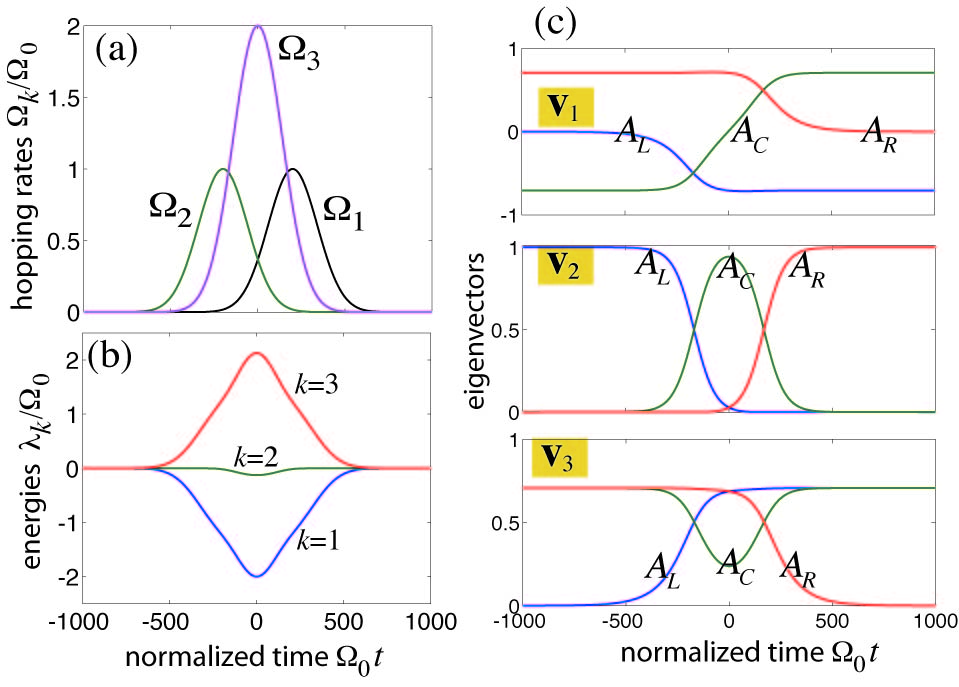}
\caption{(Color online). (a) Sequence of hopping rates and corresponding behavior of (b) adiabatic energies $\lambda_k$ and (c) eigenvectors $\mathbf{v}_k=(A_L,A_C,A_R)_k^T$ ($k=1,2,3$) of the $3 \times 3$ matrix entering in Eqs.(34). Parameter values are given in the text.}
\end{figure}

\begin{figure}[b]
\includegraphics[width=16cm]{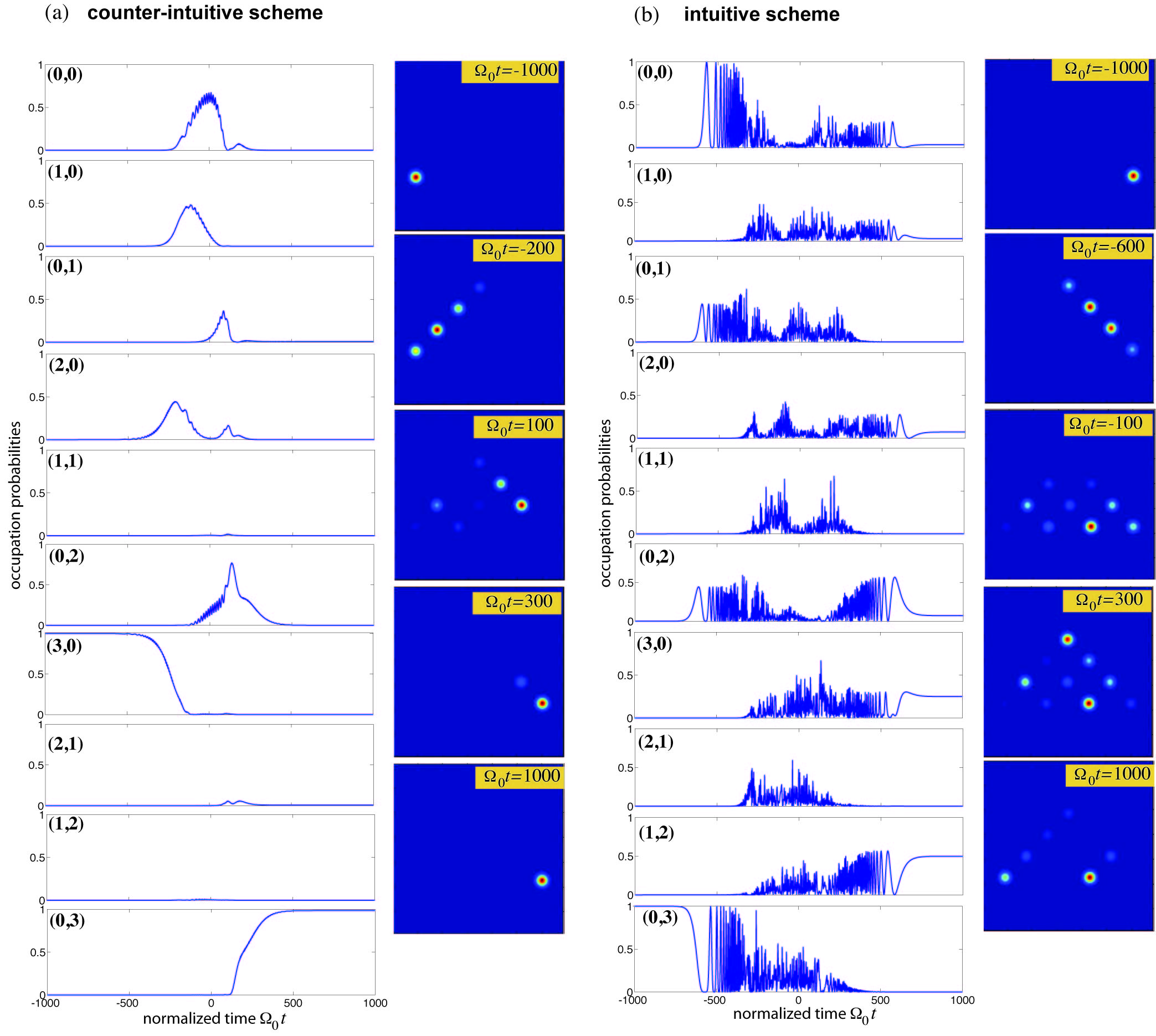}
\caption{(Color online). Numerically-computed evolution of populations $|c_{n,m}(t)|^2$ (left panels) and snapshots of site occupation at a few times (right panels) in  the triangular  lattice of Fig.3(a) with $N=3$ for initial excitation of (a) site $(3,0)$ [the counter-intuitive scheme], and (b) site $(0,3)$ [the intuitive scheme]. The dynamical evolution of tunneling rates $\Omega_1(t)$, $ \Omega_2(t)$ and $\Omega_3(t)$ is depicted in Fig.5(a).} 
\end{figure}

and $c_l=0$ for $l=2,4,5,6,8$. Under an appropriate choice of the tunneling rates $\Omega_l(t)$, the dark state can be chosen to coincide, for example, with state $|3 \rangle$ at $t \rightarrow -\infty$ and with state $|7 \rangle$ at $t \rightarrow \infty$; Fig.1(b) shows a schematic of the tunneling rates that realize such a coherent state transfer by adiabatic passage. Here, however, we follow a different approach, which enables to provide the exact analytical solutions to Eqs.(1-9) in terms of two independent three-level STIRAP processes. Most important, such an approach can be extended to different and more involved lattice geometries, as shown in the next section. The starting point of this approach is to note that the bi-dimensional lattice model (1-9) can be derived from the second-quantization Hamiltonian of six coupled bosonic oscillators
\begin{equation}
\hat{H}(t)=\hbar \Omega_1 (\hat{a}\hat{b}^{\dag}+\hat{b}\hat{a}^{\dag})+\hbar \Omega_2 (\hat{c}\hat{d}^{\dag}+\hat{d}\hat{c}^{\dag})+\hbar \Omega_3 (\hat{b}\hat{e}^{\dag}+\hat{e}\hat{b}^{\dag})+\hbar \Omega_4(\hat{
f}\hat{c}^{\dag}+\hat{c}\hat{f}^{\dag})
\end{equation}
where $\hat{a}$, $\hat{b}$, ...,$\hat{f}$ ($\hat{a}^{\dag}$, $\hat{b}^{\dag}$, ...,$\hat{f}^{\dag}$) are the annihilation (creation) operators of independent bosonic modes, satisfying the usual bosonic commutation relations. The Hamiltonian (11) conserves the total numbers of particles. Noticeably, the ninth-dimensional subspace of the two-particle sector spanned onto the vectors
\begin{eqnarray}
 |1 \rangle = \hat{a}^{\dag} \hat{f}^{\dag} | 0 \rangle , & |2 \rangle = \hat{a}^{\dag} \hat{c}^{\dag} | 0 \rangle, & |3 \rangle = \hat{a}^{\dag} \hat{d}^{\dag} | 0 \rangle   \nonumber \\
  |4 \rangle = \hat{b}^{\dag} \hat{f}^{\dag} | 0 \rangle , &  |5 \rangle = \hat{b}^{\dag} \hat{c}^{\dag} | 0 \rangle , &  | 6 \rangle = \hat{b}^{\dag} \hat{d}^{\dag} | 0 \rangle  \\
 |7 \rangle = \hat{e}^{\dag} \hat{f}^{\dag} | 0 \rangle, &  |8 \rangle = \hat{c}^{\dag} \hat{e}^{\dag} | 0 \rangle , & |9 \rangle = \hat{d}^{\dag} \hat{e}^{\dag} | 0 \rangle \nonumber
  \nonumber
\end{eqnarray}
is invariant with respect to the action of the Hamiltonian $\hat{H}(t)$. Hence, if the initial state $|\psi(t=t_i) \rangle$ at initial time $t=t_i$ belongs to this subspace, the state
vector at any time $t > t_i$, $| \psi(t) \rangle$, remains in the subspace. After setting 
\begin{equation}
|\psi(t) \rangle=\sum_{l=1}^{9} c_l(t) | l \rangle,
\end{equation}
  from the Schr\"{o}dinger equation $ i \hbar \partial_t | \psi(t) \rangle= \hat{H}(t) | \psi(t) \rangle$ it readily follows that the Fock-space amplitude probabilities $c_l(t)$ satisfy Eqs.(1-9). Hence, to compute the temporal evolution of the amplitudes $c_l(t)$ in the lattice of Fig.1(a), we can propagate the state $|\psi(t) \rangle$ and then calculate $c_l(t)$ from the scalar product 
  \begin{equation}
  c_l(t)= \langle l | \psi(t) \rangle.
  \end{equation}
   The solution $| \psi(t) \rangle $ to the Schr\"{o}dinger equation can be formally written as $| \psi(t) \rangle= \hat{U}(t,t_i) | \psi(t_i) \rangle$, i.e.
\begin{equation}
| \psi(t) \rangle=\sum_{l=1}^9 c_l(t_i) \hat{U}(t,t_i) | l \rangle
\end{equation}
 where $\hat{U}(t,t_i)$ is the unitary operator (propagator) that describes the evolution from state $|\psi(t_i) \rangle$ at initial time $t=t_i$ to the state $|\psi(t) \rangle$ at time $t=t$. The propagator $\hat{U}(t,t_i)$ satisfies the Schr\"{o}dinger equation $i \hbar (\partial \hat{U}/ \partial t)=  \hat{H}(t) \hat{U}(t,t_i)$ with $\hat{U}(t_i,t_i)=\hat{I}$ and $\hat{U}(t_i,t)=\hat{U}^{\dag}(t,t_i)$, where $\hat{I}$ is the identity operator. From Eqs.(14) and (15) one then obtains
 \begin{equation}
 c_l(t)=\sum_{m}c_m(t_i) \langle l | \hat{U}(t,t_i) |m \rangle=\sum_m \Theta_{m,l}^*(t,t_i) c_m(t_i) 
 \end{equation}
 where we have set 
 \begin{equation}
 \Theta_{m,l}(t,t_i) \equiv \langle m | \hat{U}^{\dag}(t,t_i)| l \rangle
 \end{equation}
 To evaluate the matrix elements $\Theta_{m,l}(t,t_i)$, let us note that $\hat{U}^{\dag}(t,t_i) | l \rangle$ is readily calculated by taking the expressions of $|l \rangle$, as given by Eqs.(12), with the formal substitutions $\hat{a}^{\dag} \rightarrow \hat{A}^{\dag}(t)$, $\hat{b}^{\dag} \rightarrow \hat{B}^{\dag}(t)$, $\hat{c}^{\dag} \rightarrow \hat{C}^{\dag}(t)$, ... where
 ${A}^{\dag}(t)=\hat{U}^{\dag}(t,t_i) \hat{a}^{\dag} \hat{U}(t,t_i)$, $\hat{B}^{\dag}(t)=\hat{U}^{\dag}(t,t_i) \hat{b}^{\dag} \hat{U}(t,t_i)$, $\hat{C}^{\dag}(t)=\hat{U}^{\dag}(t,t_i) \hat{c}^{\dag} \hat{U}(t,t_i)$ ... are the Heisenberg operators associated to $\hat{a}^{\dag}$, $\hat{b}^{\dag}$, $\hat{c}^{\dag}$, ... To justify this rule, let us calculate, as an example, $\hat{U}^{\dag}(t,t_i) |1 \rangle=\hat{U}^{\dag}(t,t_i) \hat{a}^{\dag} \hat{c}^{\dag} |0 \rangle$. Taking into account that $\hat{U}(t,t_i) \hat{U}^{\dag}(t,t_i)= \hat{U}^{\dag}(t,t_i) \hat{U}(t,t_i)= \hat{I}$ and $\hat{U}(t,t_i) |0 \rangle= \hat{U}^{\dag}(t,t_i) |0 \rangle=|0 \rangle$, one can write
 \begin{equation}
 \hat{U}^{\dag}(t,t_i) \hat{a}^{\dag} \hat{c}^{\dag} |0 \rangle= \hat{U}^{\dag}(t,t_i) \hat{a}^{\dag} \hat{U}(t,t_i) \hat{U}^{\dag} (t,t_i) \hat{c}^{\dag}  \hat{U}(t,t_i) \hat{U}^{\dag} (t,t_i)  |0 \rangle
 \end{equation}
 i.e.
 \begin{equation}
 \hat{U}^{\dag}(t,t_i) \hat{a}^{\dag} \hat{c}^{\dag} |0 \rangle= \hat{A}^{\dag}(t) \hat{C}^{\dag}(t) | 0 \rangle.
 \end{equation}
 The operators $\hat{A}^{\dag}(t)$,  $\hat{B}^{\dag}(t)$, $\hat{C}^{\dag}(t)$, ...  turn out to be linear combinations of the operators $\hat{a}^{\dag}$, $\hat{b}^{\dag}$, $\hat{c}^{\dag}$, ... In fact, they satisfy the Heisenberg equations of motion $i \hbar (d \hat{A}^{\dag} /dt)=-[\hat{H},\hat{A}^{\dag}]$, 
 $i \hbar (d \hat{B}^{\dag}/dt)=-[\hat{H},\hat{B}^{\dag}]$, .... with the initial condition $\hat{A}^{\dag}(t_i)=\hat{a}^{\dag}$, 
 $\hat{B}^{\dag}(t_i)=\hat{b}^{\dag}$,  .... The Heisenberg equations of motion for the six bosonic operators read explicitly 
 \begin{eqnarray}
 i \frac{d \hat{A}^{\dag}}{dt} & = &- \Omega_1(t) \hat{B}^{\dag} \nonumber \\
 i \frac{d \hat{B}^{\dag}}{dt} & = &- \Omega_1(t) \hat{A}^{\dag} -\Omega_3(t) \hat{E}^{\dag} \\
 i \frac{d \hat{E}^{\dag}}{dt} & = &- \Omega_3(t) \hat{B}^{\dag} \nonumber
 \end{eqnarray}
 for the operators $\hat{A}^{\dag}(t)$, $\hat{B}^{\dag}(t)$, $\hat{E}^{\dag}(t)$, and
 \begin{eqnarray}
 i \frac{d \hat{D}^{\dag}}{dt} & = &- \Omega_2(t) \hat{C}^{\dag} \nonumber \\
  i \frac{d \hat{C}^{\dag}}{dt} & = &- \Omega_2(t) \hat{D}^{\dag}-\Omega_4(t) \hat{F}^{\dag} \\
 i \frac{d \hat{F}^{\dag}}{dt} & = &- \Omega_4(t) \hat{C}^{\dag} \nonumber
 \end{eqnarray}
 for the operators $\hat{D}^{\dag}(t)$, $\hat{C}^{\dag}(t)$, $\hat{F}^{\dag}(t)$. Note that Eqs.(20) and (21) are formally analogous to two sets of decoupled three-level equations of STIRAP under exact resonance conditions, provided that the operators are replaced by $c$-numbers. Let us indicate by $S_{n,m}$  the $3 \times 3$ propagation matrix of the STIRAP system (20) from $t=t_i$ to some final time $t=t_f$, and similarly by  $T_{n,m}$ the propagation matrix for the STIRAP system (21).  Hence
 \begin{eqnarray}
 \hat{A}^{\dag}(t_f) & = & S_{1,1} \hat{a}^{\dag}+S_{1,2} \hat{b}^{\dag} +S_{1,3} \hat{e}^{\dag} \nonumber \\
 \hat{B}^{\dag}(t_f) & = & S_{2,1} \hat{a}^{\dag}+S_{2,2} \hat{b}^{\dag} +S_{2,3} \hat{e}^{\dag} \\
 \hat{E}^{\dag}(t_f) & = & S_{3,1} \hat{a}^{\dag}+S_{3,2} \hat{b}^{\dag} +S_{3,3} \hat{e}^{\dag} \nonumber
 \end{eqnarray}
and 
\begin{eqnarray}
 \hat{D}^{\dag}(t_f) & = & T_{1,1} \hat{d}^{\dag}+T_{1,2} \hat{c}^{\dag} +T_{1,3} \hat{f}^{\dag} \nonumber \\
 \hat{C}^{\dag}(t_f) & = & T_{2,1} \hat{d}^{\dag}+T_{2,2} \hat{c}^{\dag} +T_{2,3} \hat{f}^{\dag} \\
 \hat{F}^{\dag}(t_f) & = & T_{3,1} \hat{d}^{\dag}+T_{3,2} \hat{c}^{\dag} +T_{3,3} \hat{f}^{\dag} \nonumber .
 \end{eqnarray}
From Eqs.(16), (17), (22) and (23) the most general solution to Eqs.(1-9), from $t=t_i$ to $t=t_f$, can be then readily calculated. In particular one has
\begin{eqnarray}
c_3(t_f) & = & S_{1,1}^*T_{1,3}^*c_1(t_i)+S_{1,1}^*T_{1,2}^*c_2(t_i)+ S_{1,1}^*T_{1,1}^*c_3(t_i)+S_{1,2}^*T_{1,3}^*c_4(t_i)+ S_{1,2}^*T_{1,2}^*c_5(t_i)+S_{1,2}^*T_{1,1}^*c_6(t_i)+ \nonumber \\
& + & S_{1,3}^*T_{1,3}^*c_7(t_i)+S_{1,3}^*T_{1,2}^*c_8(t_i)+S_{1,3}^*T_{1,1}^*c_9(t_i) \\
c_7(t_f) & = &  S_{3,1}^*T_{3,3}^*c_1(t_i)+S_{3,1}^*T_{3,2}^*c_2(t_i)+ S_{3,1}^*T_{3,1}^*c_3(t_i)+S_{3,2}^*T_{3,3}^*c_4(t_i)+S_{3,2}^*T_{3,2}^*c_5(t_i)+S_{3,2}^*T_{3,1}^*c_6(t_i)+\nonumber \\
& + &S_{3,3}^*T_{3,3}^*c_7(t_i)+S_{3,3}^*T_{3,2}^*c_8(t_i)+S_{3,3}^*T_{3,1}^*c_9(t_i).
\end{eqnarray} 
We are now ready to show that exact coherent transfer between the two vertex sites $|3 \rangle$ and $|7 \rangle$ (or, similarly, between the other vertex sites $|1 \rangle$ and $|9 \rangle$) can be 
realized in the two-dimensional lattice of Fig.1(a) in the adiabatic regime under a suitable choice of the sequences, $(\Omega_{1}(t),\Omega_3(t))$ and  $(\Omega_{2}(t),\Omega_4(t))$, of the two STIRAP processes (20) and (21).  In the adiabatic limit, the expressions of the matrix coefficients $S_{n,m}$ and $T_{n,m}$ can be calculated analytically from the analysis of instantaneous eigenvectors and corresponding eigenvalues of the matrices entering in Eqs.(20) and (21). To realize adiabatic passage from the site $|3 \rangle$ to the site $|7 \rangle$, let us assume $t_i \rightarrow -\infty$, $t_f \rightarrow \infty$ with the initial condition $c_l(-\infty)=\delta_{l,3}$ (the particle is initially on site $|3 \rangle$) and let us consider a sequence for the hopping rates $\Omega_l(t)$ as schematically depicted in Fig.1(b). One then obtains
\begin{equation}
S= \left(
\begin{array}{ccc}
0 & i \sin \left( \int_{-\infty}^{\infty} dt \sqrt{\Omega_1^2+\Omega_3^2} \right)  & \cos \left( \int_{-\infty}^{\infty} dt \sqrt{\Omega_1^2+\Omega_3^2} \right) \\
0 & \cos \left( \int_{-\infty}^{\infty} dt \sqrt{\Omega_1^2+\Omega_3^2} \right)  & i \sin \left( \int_{-\infty}^{\infty} dt \sqrt{\Omega_1^2+\Omega_3^2} \right) \\
-1 & 0 & 0
\end{array}
\right)
\end{equation}
and a similar expression for the matrix $T$, where $\sqrt{\Omega_1^2+\Omega_3^2}$ is replaced by $\sqrt{\Omega_2^2+\Omega_4^2}$, i.e.
\begin{equation}
T= \left(
\begin{array}{ccc}
0 & i \sin \left( \int_{-\infty}^{\infty} dt \sqrt{\Omega_2^2+\Omega_4^2} \right)  & \cos \left( \int_{-\infty}^{\infty} dt \sqrt{\Omega_2^2+\Omega_4^2} \right) \\
0 & \cos \left( \int_{-\infty}^{\infty} dt \sqrt{\Omega_2^2+\Omega_4^2} \right)  & i \sin \left( \int_{-\infty}^{\infty} dt \sqrt{\Omega_2^2+\Omega_4^2} \right) \\
-1 & 0 & 0
\end{array}
\right)
\end{equation}
From Eqs.(24-27) it then readily follows that $c_7(+ \infty)=1$, i.e. perfect adiabatic state transfer from site $|3 \rangle$ to site $|7 \rangle$ is realized. To check the predictions of the adiabatic analysis, in Fig.2 we show an example of CTAP as obtained by direct numerical simulations of Eqs.(1-9). In the simulations, we assumed $\Omega_1(t)=\Omega_2(t)= \Omega_0  \exp \{ -[(t-\tau/2)/T_p]^2\}$, 
$\Omega_3(t)=\Omega_4(t)= \Omega_0 \exp \{ -[(t+\tau /2)/T_p]^2\}$, with parameter values $\Omega_0T_p=30$ and $\tau=T_p$.  In Fig.2(a) the initial condition is $c_{l}(0)=\delta_{l,3}$, which corresponds to the excitation of the dark state (10), an CTAP from state $|3 \rangle$ to state $| 7 \rangle$ is obtained with almost $100 \%$ fidelity. Note that, according to the form of the dark state [Eq.(10)], during the CTAP process the populated states are not only the initial ($|3 \rangle$) and final ($|7 \rangle$) sites, like in the ordinary three-well CTAP in a linear chain, but also the other two vertices  $|1 \rangle$ and $|9 \rangle$ of the square lattice. Indeed, as discussed in the beginning of this section, the CTAP scheme of Fig.2(a) can be viewed as the superposition of two independent one-dimensional STIRAP processes in the horizontal and vertical directions, and this explains why all four sites at the vertices of the square lattice are excited during the adiabatic passage.
The numerical results in Fig.2(b)  correspond to the 'wrong' initial condition $c_{l}(0)=\delta_{l,7}$ (the particle is initially on site $|7 \rangle$ rather than on site $|3 \rangle$). In this case the initial state does not coincide with the dark state (10), and the adiabatic passage to the final state $|3 \rangle$ does not occur.\par
As a final note, we mention that the STIRAP scheme discussed so far for the rectangular lattice of Fig.1(a) can be implemented, for example, in an array of $3 \times 3$ optical waveguides with controlled separation distances, generalizing the optical set ups discussed in Refs.\cite{A13,A14}.   

\section{CTAP in a triangular lattice}
As a second example of exactly-solvable CTAP problem in a bi-dimensional lattice, let us consider the hopping motion of a quantum particle on the triangular lattice of Fig.3(a), composed by $(N+1)(N+2)/2$ sites ($N=0,1,2,...$). The hopping rates among the lattice sites are engineered as shown in the figure and depend on three independent parameters $\Omega_1(t)$, $\Omega_2(t)$ and $\Omega_3(t)$. We will show that, under an appropriate tuning of the amplitudes $\Omega_1(t)$, $\Omega_2(t)$ and $\Omega_3(t)$, CTAP can be realized between the two vertex sites $(N,0)$ and $(0,N)$ of the triangular lattice. We note that the triangular lattice model of Fig.3(a) can be regarded as a generalization of the half square lattice previously studied by Bradly and coworkers in Ref.\cite{A20tris}, which is obtained from our analysis in the limiting case $\Omega_3(t)=0$.
In the tight-binding and nearest-neighbor approximations, the amplitude probabilities $c_{n,m}(t)$ to find the particle at site $(n,m)$ of the triangular lattice evolve according to the coupled equations
\begin{eqnarray}
i \frac{dc_{n,m}}{dt} & = & \Omega_1(t) \sqrt{n(N+1-n-m)} c_{n-1,m}+ \Omega_1(t) \sqrt{(n+1)(N-n-m)}c_{n+1,m}  \nonumber \\
& + & \Omega_2(t)  \sqrt{m(N+1-n-m)} c_{n,m-1}+\Omega_2(t)  \sqrt{(m+1)(N-n-m)} c_{n,m+1} \\
& + & \Omega_3(t) \sqrt{n(m+1)}c_{n-1,m+1}+ \Omega_3(t) \sqrt{m(n+1)}c_{n+1,m-1} \nonumber
\end{eqnarray}
where $n,m \geq 0$ and $n+m \leq N$. For $\Omega_3(t)=0$, this lattice model can be regarded as a two-dimensional extension of the so-called Krawtchouk quantum chain \cite{K1,K2,K3}, which in the one-dimensional case can be mapped into a non-interacting bosonic junction \cite{Longhi14}. The bi-dimensional extension of the Krawtchouk quantum chain [Eq.(28)] with $\Omega_3(t)=0$ basically describes the dynamics of a single particle on a half square lattice of size $N$, cut along one of the main diagonals. This problem  can be mapped onto the dynamics of $N$ non-interacting bosonic particles in a triple well potential, as previously shown in Ref.\cite{A20tris}. Here we extend the analysis by allowing for a non-vanishing value of $\Omega_3(t)$, i.e. by considering an effective triangular (rather than half-square) lattice. The second-quantization Hamiltonian that describes the hopping dynamics of non-interacting bosons in a triple well potential [see Fig.3(b)] reads
\begin{equation}
\hat{H}(t)=\hbar \Omega_1(t) \left( \hat{a}^{\dag}_L \hat{a}_C+\hat{a}^{\dag}_C \hat{a}_L \right)+ \hbar \Omega_2(t) \left( \hat{a}^{\dag}_R \hat{a}_C+\hat{a}^{\dag}_C \hat{a}_R \right)+\hbar \Omega_3(t) \left( \hat{a}^{\dag}_L \hat{a}_R+\hat{a}^{\dag}_R \hat{a}_L \right)
\end{equation}
where $L$ (left), $C$ (central) and $R$ (right) are the three well sites, $\hat{a}^{\dag}_K$ and $\hat{a}_K$ ($K=L,C,R$) are the bosonic creation and annihilation
operators in the three wells, and  $\Omega_1(t)$, $\Omega_2(t)$, $\Omega_3(t)$  are the hopping (tunneling) rates between the wells $L-C$, $R-C$ and $L-R$, respectively. The total number of bosons $\hat{N} =\hat{a}^{\dag}_L \hat{a}_L+\hat{a}^{\dag}_C \hat{a}_C+\hat{a}^{\dag}_R \hat{a}_R$ is a conserved quantity and the dimension of the
Hilbert space is $(N+1)(N + 2)/2$. In the $N$-particle sector of Hilbert space, the state vector $| \psi(t) \rangle$ of the bosonic field can be expanded as
\begin{equation}
| \psi(t) \rangle=\sum_{n,m} c_{n,m}(t) |n,m \rangle 
\end{equation}
where the indices $n,m$ vary from $0$ to $N$, with $n+m \leq N$. In Eq.(30), the Fock state $|n,m \rangle$ is defined by 
\begin{equation}
|n,m \rangle = |n \rangle_L | m \rangle_R | N-n-m \rangle_C=\frac{1} {\sqrt{n!m ! (N-n-m)!}} \hat{a}^{\dag \; n}_L \hat{a}^{\dag \; m}_R \hat{a}^{\dag \; N-n-m}_C |0 \rangle
\end{equation}
 and corresponds to $n$ bosons trapped in the left well, $m$ bosons in the right well, and the remaining $(N-n-m)$ bosons in the central well. 
 Substitution of Eq.(30) into the Schr\"{o}dinger equation $ i \hbar \partial_t | \psi(t) \rangle= \hat{H}(t) | \psi(t) \rangle$ yields for the amplitude probabilities $c_{n,m}(t)$ the evolution equations as given by Eq.(28). The equivalence between the non-interacting bosonic triple well model of Fig.3(b) and the  the bi-dimensional lattice of Fig.3(a) can be exploited to derive an analytical solution for the temporal evolution of amplitude probabilities $c_{n.m}(t)$ of Eq.(28).  To this aim, we follow a procedure similar to the one described in the previous section, which yields [compare with Eqs.(16) and (17)]
 \begin{equation}
 c_{n,m}(t)=\sum_{p,q} c_{p,q}(t_i) \langle n,m| \hat{U}(t,t_i) | p,q \rangle =\sum_{p,q} \Theta^{*}_{p,q; n,m}(t,t_i) c_{p,q}(t_i)
 \end{equation}
  where $\hat{U}(t,t_i)$ is the propagator associated to the Hamiltonian $\hat{H}(t)$ from the initial time $t_i$ to time $t$, and where we have set
 \begin{equation}
\Theta_{p,q; n,m}(t,t_i)= \langle p,q | \hat{U}^{\dag}(t,t_i)| n,m \rangle.
 \end{equation}
 The scalar products in Eq.(33) can be readily computed after expressing the state $\hat{U}^{\dag}(t,t_i)| n,m \rangle$ as a linear combination of states $|l,s \rangle$ by noting that 
 $\hat{U}^{\dag}(t,t_i)| n,m \rangle$ is formally expressed by Eq.(31) after replacement of the operators $\hat{a}^{\dag}_K$ ($K=L,C,R$) with their evolved Heisenberg operators $\hat{A}^{\dag}_K(t)$, where ${A}^{\dag}_K(t)=\hat{U}^{\dag}(t,t_i) \hat{a}^{\dag}_K \hat{U}(t,t_i)$. The operators ${A}^{\dag}_L(t)$, ${A}^{\dag}_C(t)$, ${A}^{\dag}_R(t)$ evolve according to the Heisenberg equations, which read explicitly
 \begin{eqnarray}
 i \frac{d \hat{A}^{\dag}_L}{dt} & = & - \Omega_1(t) \hat{A}^{\dag}_C-\Omega_3(t) \hat{A}^{\dag}_R  \nonumber \\
 i \frac{d \hat{A}^{\dag}_C}{dt} & = & - \Omega_1(t) \hat{A}^{\dag}_L - \Omega_2(t) \hat{A}^{\dag}_ R  \\
 i \frac{d \hat{A}^{\dag}_R}{dt} & = & - \Omega_2(t) \hat{A}^{\dag}_C - \Omega_3(t) \hat{A}^{\dag}_L \nonumber 
 \end{eqnarray}
 Note that Eqs.(34) are formally analogous to the STIRAP equations of a three-level system under exact resonance, but with next-nearest hopping between wells $R$ and $L$. The three-well STIRAP scheme with $\Omega_3 \neq 0$ has been recently introduced and discussed by Menchon-Enrich and collaborators in Ref.\cite{A20}.\\  
 Indicating by $S$ the $3 \times 3$ propagation matrix of the STIRAP system (34) from $t=t_i$ to some final time $t=t_f$, one can write
  \begin{eqnarray}
 \hat{A}^{\dag}_L(t_f) & = & S_{1,1} \hat{a}^{\dag}_L+S_{1,2} \hat{a}^{\dag}_C +S_{1,3} \hat{a}^{\dag}_R \nonumber \\
 \hat{A}^{\dag}_C(t_f) & = & S_{2,1} \hat{a}^{\dag}_L+S_{2,2} \hat{a}^{\dag}_C +S_{2,3} \hat{a}^{\dag}_R \\
 \hat{A}^{\dag}_R(t_f) & = & S_{3,1} \hat{a}^{\dag}_L+S_{3,2} \hat{a}^{\dag}_C +S_{3,3} \hat{a}^{\dag}_R \nonumber
 \end{eqnarray}
 Using Eqs.(31) and (35), the expressions of the coefficients $\Theta_{p,q;n,m}(t_f,t_i)$ can be thus determined in a closed form
 \begin{eqnarray}
 \Theta_{p,q;n,m} & = & \frac{1}{\sqrt{n!m!p!q!(N-n-m)!(N-p-q)!}}  \\
 & \times & \langle 0 |  \hat{a}^{ p}_{L} \hat{a}^{ q}_R \hat{a}^{ N-p-q}_C (S_{11}\hat{a}^{\dag}_L+S_{12}\hat{a}^{\dag}_C+S_{13}\hat{a}^{\dag}_R)^{n} (S_{31}\hat{a}^{\dag}_L+S_{32}\hat{a}^{\dag}_C+S_{33}\hat{a}^{\dag}_R)^{m} (S_{21}\hat{a}^{\dag}_L+S_{22}\hat{a}^{\dag}_C+S_{23}\hat{a}^{\dag}_R)^{N-n-m}
   |0 \rangle  \nonumber
 \end{eqnarray}
 We are now ready to show that CTAP can be realized between the outer sites $(0,N)$ and $(N,0)$ in the triangular lattice of Fig.3(a). Let us assume that at initial time $t_i \rightarrow -\infty$ the particle occupies the site $(N,0)$, i.e. $c_{p,q}(t_i)=\delta_{p,N} \delta_{q,0}$, and let us tune the hopping rates $\Omega_1(t)$, $\Omega_2(t)$ and $\Omega_3(t)$ in the counter-intuitive scheme shown in Fig.3(c). At final time $t_f $ one has $c_{n,m}(t_f)=\Theta_{N,0; n,m}^*(t_f, -\infty)$, i.e.
\begin{equation}
c_{n,m}(t_f)=\sqrt{\frac{N!}{n! m! (N-n-m)!}} S_{11}^{* \; n}  S_{31}^{* \; m}  S_{21}^{* \; N-n-m}
\end{equation} 
where we used Eq.(36).
 In the adiabatic approximation the coefficients of the STIRAP matrix $S$ can be calculated in a closed form.\par
 Let us first consider the half-square lattice limit $\Omega_3(t)=0$, which was previously investigated in Ref.\cite{A20tris}. In this case the STIRAP  matrix at $t_f \rightarrow \infty$ reads
 \begin{equation}
S= \left(
\begin{array}{ccc}
0 & i \sin \left( \int_{-\infty}^{\infty} dt \sqrt{\Omega_1^2+\Omega_2^2} \right)  & \cos \left( \int_{-\infty}^{\infty} dt \sqrt{\Omega_1^2+\Omega_2^2} \right) \\
0 & \cos \left( \int_{-\infty}^{\infty} dt \sqrt{\Omega_1^2+\Omega_2^2} \right)  & i \sin \left( \int_{-\infty}^{\infty} dt \sqrt{\Omega_1^2+\Omega_2^2} \right) \\
-1 & 0 & 0
\end{array}
\right)
\end{equation}
 From Eqs.(37) and (38) it then readily follows that $c_{n,m}(\infty)$ is non-vanishing solely for $n=0$ and $m=N$, i.e. at final time the particle occupies the lattice site $(0,N)$. Hence CTAP from state $(N,0)$ to state $(0,N)$ in the lattice of Fig.3(a) is realized. In the second-quantization picture of Fig.3(b), the CTAP is clearly explained as 
 an ordinary CTAP in a triple well system for $N$ non-interacting particles. In the original half-square lattice of Fig.3(a), the CTAP process could be explained on the basis of a kind of multilevel STIRAP scheme in a system with an odd number of levels \cite{multiSTIRAP}. The dark state of the system corresponds to $c_{n,m}=0$ for $n+m < N$ and
 \begin{equation}
 c_{N-l,l}=(-1)^l \mathcal{N} \left( \frac{\Omega_1}{\Omega_2} \right)^{l} \sqrt{\frac{N!}{l! (N-l)!}}
 \end{equation}
 ($l=0,1,2,...,N$), where $\mathcal{N}$ is a normalization constant, given by
 \begin{equation}
 \frac{1}{\mathcal{N}}=\sqrt{\sum_{l=0}^{N} 
 \left(
 \begin{array}{c}
  N \\
  l
 \end{array} 
 \right) \left( \frac{\Omega_1}{\Omega_2} \right)^{2l}}=\left[ 1+ \left( \frac{\Omega_1}{\Omega_2}\right)^2 \right]^{N/2}.
 \end{equation}
 Note that, for $(\Omega_1 / \Omega_2) \rightarrow 0$, the dark state has only one non-vanishing element, namely $c_{N,0}$, whereas for $(\Omega_1 / \Omega_2) \rightarrow \infty$ the only non-vanishing element is $c_{0,N}$. The adiabatic transfer thus results from the excitation of this dark state and in its adiabatic evolution for the sequence of $\Omega_{1,2}(t)$ shown in Fig.3(c) (see also Ref.\cite{A20tris}).  Therefore, the CTAP scheme in the $\Omega_3=0$ limit basically  reduces to a multilevel STIRAP in the linear chain of sites $(N,0)$, $(N-1,0)$, $(N-1,1)$, $(N-2,1)$, $(N-2,2)$,  ..., $(0,N-1)$, $(0,N)$ at the bottom edge of the half-square lattice. To check the predictions of the adiabatic analysis, in Fig.4 we show an example of CTAP as obtained by direct numerical simulations of Eqs.(28) with $\Omega_3(t)=0$ in a half-square lattice with $N=3$, comprising $(N+1)(N+2)/2=10$ lattice sites. In the simulations, we assumed $\Omega_1(t)= \Omega_0 \exp \{ -[(t-\tau /2)/T_p]^2\}$ and
$\Omega_2(t)= \Omega_0 \exp \{ -[(t+\tau /2 )/T_p]^2\}$, with parameter values $\Omega_0T_p=30$ and $\tau=T_p$ (see the upper inset in Fig.4).  In Fig.4(a) the initial condition is $c_{n,m}(0)=\delta_{n,N} \delta_{m,0}$ [the particle is initially in the state $(N,0)$], and CTAP to the state $(0,N)$ is clearly observed. Note that in the evolution only the states $(n,m)$ of the lattice with $n+m=N=3$ (i.e. the last row of the lattice) are basically excited, which is in agreement with the form of the dark state for the half-square  lattice [see Eq.(39)]. Conversely, in Fig.4(b) the numerical results correspond to the 'wrong' initial condition $(0,N)$. In this case CTAP to the final state $(N,0)$ is not observed for the chosen sequence of $\Omega_{1,2}(t)$. \par
Let us now consider the more general case $\Omega_3(t) \neq 0$. The possibility to realize STIRAP in the three-well system  of Fig.3(b) with next-nearest interaction (i.e. for $ \Omega_3 \neq 0$) was discussed in details in Ref.\cite{A20} for the single particle case. The eigenenergies $\lambda_k$ and corresponding adiabatic eigenstates $\mathbf{v}=(A_L,A_C,A_R)^T$ of Eqs.(34) with $\Omega_3 \neq 0$ are given by \cite{A20}
\begin{equation}
\lambda_k=2 \sqrt{-\frac{p}{3}} \cos \left[ \frac{1}{3}  {\rm acos} \left( \frac{3q}{2p} \sqrt{-\frac{3}{p}} \right)  +k \frac{2 \pi}{3} \right]
\end{equation}
\begin{equation}
\mathbf{v}_k \equiv = \left(
\begin{array}{c}
A_L \\
A_C \\
A_R
\end{array}
\right)_k=
\left(
\begin{array}{c}
 a_k / \mathcal{N}_k \\
 b_k / \mathcal{N}_k\\
 -c_k / \mathcal{N}_k
\end{array}
\right)
\end{equation}
$(k=1,2,3$), where we have set 
\begin{equation}
p=-(\Omega_1^2+\Omega_2^2+\Omega_3^2) \; , \;\; q= 2 \Omega_1 \Omega_2 \Omega_3
\end{equation}
\begin{equation}
a_k=\Omega_2+\frac{\lambda_k \Omega_3}{\Omega_1} \, , \; \;  b_k=\Omega_3+\frac{\lambda_k \Omega_2}{\Omega_1} \;,\;\; c_k=\Omega_1-\frac{\lambda_k^2}{\Omega_1}
\end{equation}
and $\mathcal{N}_k=(a_k^2+b_k^2+c_k^2)^{1/2}$. For a sequence of hopping rates schematically shown in Fig.3(c), the adiabatic eigenstate $\mathbf{v}_2$ reduces to $(1,0,0)$ at $t \rightarrow -\infty$ and to $(0,0,1)$ at $t \rightarrow \infty$. CTAP can be thus achieved in the adiabatic limit, provided that the eigenenergy $\lambda_2$ does not cross none of the other two energies $\lambda_1$ and $\lambda_3$. Such a condition set some constraints on the strength of $\Omega_3$, which have been discussed in Ref.\cite{A20}. An example of hopping rate sequence that avoids energy crossing is shown, for instance, in Fig.5(a), together with the evolution of the adiabatic energies and corresponding eigenvectors [Figs.5(b) and (c)]. 
The hopping rates shown in the figure are Gaussian-shaped and given by $\Omega_1(t)= \Omega_0 \exp \{ -[(t-\tau /2)/T_p]^2\}$, $\Omega_2(t)= \Omega_0 \exp \{ -[(t+\tau/2)/T_p]^2\}$, and $\Omega_3(t)= \sigma \Omega_0 \exp [ -(t/T_p)^2 ]$, with parameter values $\Omega_0 T_p=100$, $\tau=2 T_p$ and $\sigma=2$. Note that for such a sequence of hopping rates there is not any energy crossing and thus, if the three well system of Fig.3(b) is initially prepared in one of its eigenstates, it remains in the eigenstate under slow (adiabatic) change of the hopping rates. Let us consider now the hopping dynamics on the triangular lattice of Fig.3(a), and let us assume that the particle at initial time $t_i \rightarrow -\infty$ occupies the state $(N,0)$. At final time $t_f $, the occupation amplitudes of the various lattice sites are given again by Eq.(37), where the matrix coefficients $S_{1,2}$, $S_{2,1}$ and $S_{3,1}$ now reads
\begin{equation}
\left( 
\begin{array}{c}
S_{1,1}(t_f)\\
S_{2,1}(t_f)\\
S_{3,1}(t_f)
\end{array}
\right)=
\left( 
\begin{array}{c}
A_{L}(t_f)\\
A_{C}(t_f)\\
A_{R}(t_f)
\end{array}
\right)_2 \exp 	\left(-i \int_{-\infty}^{t_f} dt \lambda_2(t) \right)
\end{equation}
i.e. they are given by the elements of the eigenvector $\mathbf{v}_2$, multiplied by the corresponding adiabatic phase. For $t_f \rightarrow \infty$, in the adiabatic limit and provided that energy crossing is avoided one thus obtains $S_{11}(t_f)=S_{21}(t_f)=0$ and $|S_{3,1}(t_f)|=1$, so that from Eq.(37) it follows that $|c_{n,m}(t_f)|=\delta_{0,N}$: this means that particle has been  transferred from the site $(N,0)$  to the site $(0,N)$ of the triangular lattice with $100 \%$ fidelity. We checked the prediction of the adiabatic analysis by direct numerical simulations of  Eqs.(28) in a triangular lattice with $N=3$, assuming the pulse sequence shown in Fig.5(a). The numerical results are given in Fig.6, corresponding to either initial excitation of the site $(N,0)$ [the counter-intuitive scheme, Fig.6(a)] and of the state $(0,N)$   [the intuitive scheme, Fig.6(b)]. Note that in the former case CTAP to the final state $(0,N)$ is observed, according to the adiabatic analysis. It is worth comparing the CTAP process in the half-square lattice of Fig.4(a), corresponding to $\Omega_3=0$, with the CTAP process in the triangular lattice of Fig.6(a). While in the former case the two-dimensional CTAP process basically reduces to a kind of multilevel STIRAP process in a linear chain involving the bottom edge sites (as previously discussed), in the latter case this is not the case and many other sites of the triangular lattice are excited during the adiabatic process [see Fig.6(a)]. The reason thereof is that, for $\Omega_3 \neq 0$ the single-particle adiabatic state $\mathbf{v}_2$ of the three-well system of Fig.3(b) populates the central well $C$, in addition to the left $L$ and right $R$ wells [see Fig.5(c)]. Hence the CTAP in the bi-dimensional triangular lattice is a rather nontrivial effect, that cannot be reduced to neither the composition of two one-dimensional CTAP processes (like in the rectangular lattice discussed in Sec.2) nor to a one-dimensional multilevel STIRAP (like in the half-square lattice). \par

 Finally, we would like to briefly mention that a possible physical implementation of the triangular lattice of Fig.3(a) could be realized in a {\it linear} spatial chain of $M=(N+1)(N+2)/2$ trapped ions, where the effective ion-ion couplings in the chain can be rather arbitrarily controlled and tuned by an optical modulation scheme using a suitable number of independent optical beams \cite{ions}. The basic idea is that the hopping dynamics in the bi-dimensional triangular lattice of Fig.3(a) can be mapped into the dynamics of spin excitation in a linear chain of $M$ atoms with non-nearest neighborhood couplings that reproduce the hopping scheme in the original bi-dimensional lattice.
 The trapped ion system is described by the Hamiltonian
 \begin{equation}
 \hat{H}=\sum_{i,j=1 \; , \; i<j}^M J_{i,j} \sigma_{x}^{(i)} \sigma_{x}^{(j)}
 \end{equation}
where the spin operator $\sigma_x^{(i)}$ refers to the effective spin-1/2 system within each atom (represented, for
example, by a pair of hyperfine ground states separated by frequency $\omega_s$), and  $J_{i,j}$ is the spin-spin coupling strength between atoms $i$ and $j$. 
 For example, to simulate the triangular lattice of Fig.3(a) with $N=3$, a linear chain of $M=10$ atoms is required. If the sites $l=1,2,3,...,10$ in the linear chain are mapped into the sites $(0,0)$, $(1,0)$, (0,1)$, $(2,0)$, $(1,1)$, (0,2)$, $(3,0)$, $(2,1)$, $(1,2)$, $(0,3)$ of the original triangular lattice of Fig.3(a), respectively, the required Ising coupling matrix $J_{i, j}$ is given by
 \begin{equation}
 J=\left( 
 \begin{array}{cccccccccc}
 0 & \sqrt{3} \Omega_1 & \sqrt{3} \Omega_2 & 0 & 0 & 0 & 0 & 0 & 0 & 0 \\
 \sqrt{3} \Omega_1 & 0 & \Omega_3 & 2 \Omega_1 & \sqrt{2} \Omega_2 & 0 & 0 & 0 & 0 & 0 \\
 \sqrt{3} \Omega_2 & \Omega_3 & 0 & 0 & \sqrt{2} \Omega_1 & 2 \Omega_2 & 0 & 0 & 0 & 0 \\
 0 & 2 \Omega_1 & 0 & 0 & \sqrt{2} \Omega_3 & 0 & \sqrt{3} \Omega_1 & \Omega_2& 0 & 0 \\
 0 & \sqrt{2} \Omega_2 & \sqrt{2} \Omega_1 & \sqrt{2} \Omega_3 & 0 & \sqrt{2} \Omega_3 & 0 & \sqrt{2} \Omega_1 & \sqrt{2} \Omega_2 & 0\\
 0 & 0 & 2 \Omega_2 & 0 & \sqrt{2} \Omega_3 & 0 & 0 & 0 &  \Omega_1 & \sqrt{3} \Omega_2 \\
 0 & 0 & 0 & \sqrt{3} \Omega_1 & 0 & 0 & 0 & \sqrt{3} \Omega_3 & 0 & 0 \\
 0 & 0 & 0 & \Omega_2 & \sqrt{2} \Omega_1 & 0 & \sqrt{3} \Omega_3 & 0 & 2 \Omega_3 & 0\\
 0 & 0 & 0 & 0 & \sqrt{2} \Omega_2 & \Omega_1 & 0 & 2 \Omega_3 & 0 & \sqrt{3} \Omega_3 \\
 0 & 0 & 0 & 0 & 0 & \sqrt{3} \Omega_2 & 0 & 0 & \sqrt{3} \Omega_3 & 0\\
 \end{array}
 \right)
 \end{equation}
 The spins in the linear array can be coherently manipulated using a pair of counter-propagating laser beams, which
 drive stimulated Raman transitions between the spin states while also coupling off-resonantly to the collective motion of the atomic chain. 
To generate an arbitrary Ising coupling matrix $J_{i, j}$, the technique described in Ref.\cite{ions} can be employed, in which $M$ spectral beatnote detunings  to the
Raman beams are introduced, one near each motional mode with a unique pattern of spectral components on each ion. Indicating by $\Omega_{i,n}$ the Rabi frequency matrix of spectral component $n$ at ion $i$, the Ising coupling matrix is given by 
\begin{equation}
J_{i,j}=\sum_{n=1}^M \Omega_{i,n} \Omega_{j,n} F_{i,j,n},
\end{equation}
 where  $F_{i,j,n}$ characterizes the response of Ising coupling $J_{i,j}$ to spectral component $n$; the explicit form of $F_{i,j,n}$ is given by Eq.(3) of Ref.\cite{ions}. In this way, one has $M \times M$ free control parameters (the Rabi frequency spectral components addressing each ion in the chain) that at each time can be tailored to achieve the desired $M(M-1)/2$ independent elements of the (symmetric) Ising matrix $J_{i,j}$. As discussed in Ref.\cite{ions}, the determination of the Rabi matrix $\Omega_{i,n}$ from a desired Ising matrix $J_{i,j}$ can be done using a constrained
nonlinear optimization method that minimizes the total laser beam intensity; such a detailed analysis, however, goes beyond the scope of the present work.


\section{Conclusion and discussion}

Coherent tunneling by adiabatic passage is a powerful and robust technique to transport quantum states in space. Most of CTAP schemes studied so far are mainly limited to one-dimensional (ore effectively one-dimensional) geometries, whereas few works have considered CTAP models in the two-dimensional case. In this paper we have theoretically introduced two exactly-solvable bi-dimensional lattice models where CTAP can be exactly realized.
The former model provides a relatively simple extension of the ordinary three-state CTAP scheme to a $3 \times 3$ bi-dimensional lattice, whereas the second model considers a triangular lattice geometry which is nontrivially related to the ordinary three-level CTAP scheme. In both models, the hopping dynamics of the quantum particle on the bi-dimensional lattice can be described by second-quantization Hamiltonians of bosonic fields, which are quadratic in the field operators. The Heisenberg equations of motion of the bosonic operators basically describe a three-level STIRAP process (or independent three-level STIRAP processes), which ensures perfect CTAP in the original bi-dimensional lattice between suitable states of the lattice. It is envisaged that our approach could be extended to find novel and non-trivial CTAP schemes in two- (or multi-) dimensional lattices, with potential interest to coherent transport in space of matter or classical waves in engineered lattices.

\end{document}